\documentclass[12pt,preprint]{aastex}
\usepackage{psfig}

\newcommand\mzon   {M$_{\odot}$}
\newcommand\pp     {$\pm$}
\newcommand\pers   {s$^{-1}$}

\newcommand\Lunit   {erg s$^{-1}$}
\newcommand\funit   {erg cm$^{-2}$ s$^{-1}$}

\righthead{XMM-Newton observations of KS 1731--260 in quiescence}
\slugcomment{Accepted for publication in ApJ Letters: May 22, 2002}

\begin{document}

\title{{\itshape XMM-Newton} observations of the neutron star X-ray
transient KS 1731--260 in quiescence}

\author{Rudy Wijnands\altaffilmark{1,2}, Matteo
Guainazzi\altaffilmark{3}, Michiel van der Klis\altaffilmark{4},
Mariano M\'endez\altaffilmark{5}}

\altaffiltext{1}{Center for Space Research, Massachusetts Institute of
Technology, 77 Massachusetts Avenue, Cambridge, MA 02139-4307, USA;
rudy@space.mit.edu}

\altaffiltext{2}{Chandra Fellow}

\altaffiltext{3}{XMM-Newton Science Operation Center, VILSPA-ESA,
Apartado 50727, E-28080 Madrid, Spain}

\altaffiltext{4}{Astronomical Institute ``Anton Pannekoek'',
University of Amsterdam, Kruislaan 403, NL-1098 SJ Amsterdam, The
Netherlands}

\altaffiltext{5}{SRON, National Institute for Space Research,
Sorbonnelaan 2, 3584 CA, Utrecht, The Netherlands}

\begin{abstract}

We report on {\it XMM-Newton} observations performed on 2001 September
13--14 of the neutron star X-ray transient KS 1731--260 in
quiescence. The source was detected at an unabsorbed 0.5--10 keV flux
of only $4 - 8 \times10^{-14}$ \funit, depending on the model used to
fit the data, which for a distance of 7 kpc implies a 0.5--10 keV
X-ray luminosity of approximately $2 - 5\times10^{32}$ \Lunit. The
September 2001 quiescent flux of KS 1731--260 is lower than that
observed during the {\it Chandra} observation in March 2001. In the
cooling neutron star model for the quiescent X-ray emission of neutron
star X-ray transients, this decrease in the quiescent flux implies
that the crust of the neutron star in KS 1731--260 cooled down rapidly
between the two epochs, indicating that the crust has a high
conductivity.  Furthermore, enhanced cooling in the neutron star core
is also favored by our results.

\end{abstract}


\section{Introduction \label{section:intro}}

X-ray transients are characterized by long episodes (years to decades)
of very low X-ray luminosities ($10^{30-34}$ \Lunit) with occasional
short (weeks to months) outbursts during which they can be detected at
luminosities of $10^{36-39}$ \Lunit~(e.g., Chen, Shrader, \& Livio
1997). The huge increase in luminosity is thought to be due to a
correspondingly large increase in the mass accretion rate onto the
compact object in those systems, although the exact mechanisms behind
the outbursts are not fully understood (Lasota 2001).  Similarly, the
exact origin of the quiescent X-ray emission remains elusive. For
those systems harboring a neutron star, it has been argued (e.g.,
Campana et al. 1998b; Brown, Bildsten, \& Rutledge 1998) that the
observed emission below a few keV originates from the neutron star
surface: the neutron star core is heated by the nuclear reactions
occurring deep in the crust when the star is accreting and this heat
is released as thermal emission during quiescence.  The emission above
a few keV (as observed in several systems; e.g., Asai et al. 1996,
1998; Campana et al. 1998a, 2000) cannot be explained by this
model. Models proposed for this component include residual accretion
either onto the neutron star surface or down to its magnetospheric
radius, or the radio pulsar mechanism (e.g., Campana et al. 1998b;
Campana \& Stella 2000).

A sub-class of transients are characterized by very long accretion
episodes of years to decades instead of weeks to months. Recently, one
of those systems (KS 1731--260) suddenly turned off after having
actively accreted for over 12.5 years. A {\it Chandra} observation
taken a few months after this transition showed the source at a
0.5--10 keV luminosity of $\sim10^{33}$ \Lunit~(Wijnands et al. 2001),
assuming a distance of 7 kpc (Muno et al. 2000). If the cooling
neutron star model is responsible for the quiescent emission in this
system, then it should be in quiescence between outbursts for $>$1000
years, assuming all outbursts are similar to the one observed and
standard cooling processes (e.g., modified Urca; Colpi et al. 2001;
Ushomirsky \& Rutledge 2001) occur in the neutron star core.  However,
Rutledge et al. (2002) argued that for systems like KS 1731--260, the
long accreting episodes will heat the crust to high temperatures and
it might take years to decades for the crust to come into thermal
equilibrium with the core. Until this happens, the quiescent emission
will be dominated by the thermal state of the crust and not that of
the core. Rutledge et al. (2002) calculated crust cooling tracks for
this source assuming different scenarios of the microphysics involved
(the heat conductivity of the crust; standard vs. ``enhanced'' core
cooling).

Burderi et al. (2002) reported on a {\it BeppoSAX} observation of KS
1731--260 performed a few weeks before the {\it Chandra}
observation. They detected KS 1731--260 at a luminosity of at most
$\sim10^{33}$ \Lunit.  In addition to the cooling neutron star model,
they discussed several alternative explanations for the observed
quiescent emission (such as residual accretion or the onset of the
radio pulsar mechanism). By considering those alternative models, they
were able to set an upper limit of $1-4 \times 10^9$ Gauss on the
magnetic field strength of the neutron star in KS 1731--260.  Here we
report on {\it XMM-Newton} observations of KS 1731--260 taken
approximately half a year after the {\it Chandra} and {\it BeppoSAX}
observations. With these {\it XMM-Newton} observations we are able to
study the time evolution of the quiescent emission.

\section{Observation, analysis, and results}

We have analyzed {\it XMM-Newton} observations of KS 1731--260
performed on 13 September 2001 01:54--09:01 UTC and 13--14 September
2001 22:43--05:58 UTC. All instruments were active; here we only
discuss the data as obtained with the three European Photon Imaging
Camera (EPIC) instruments (due to the very low flux of the source, it
was not detected in the RGS instrument). The two EPIC MOS cameras and
the EPIC pn camera operated in full window mode with the thin optical
blocking filter.  To analyze the data, we used the Science Analysis
System (SAS\footnote{See http://xmm.vilspa.esa.es/user/sas\_top.html};
version 5.2). We used the calibrated pipeline product data to extract
images, light curves, and spectra using the tools available in SAS.
Several background flares occurred during our observations, which were
filtered out before analyzing the data to minimize the effects of
those strong background flares on the quality of the X-ray spectra; we
did not use those data during which the count rate exceeded 7 counts
\pers~for the MOS cameras (using time bins of 10 seconds) and 20
counts \pers~for the pn camera (also using 10 seconds time
bins). These criteria resulted in a total good time of $\sim$23 ksec
for the pn camera and $\sim$33 ksec for both MOS cameras. No
difference in the count rates between the two {\it XMM-Newton}
observations was observed, and, therefore, we combined the data of
both observations to increase our sensitivity.

\begin{figure}
\begin{center}
\begin{tabular}{c}
\psfig{figure=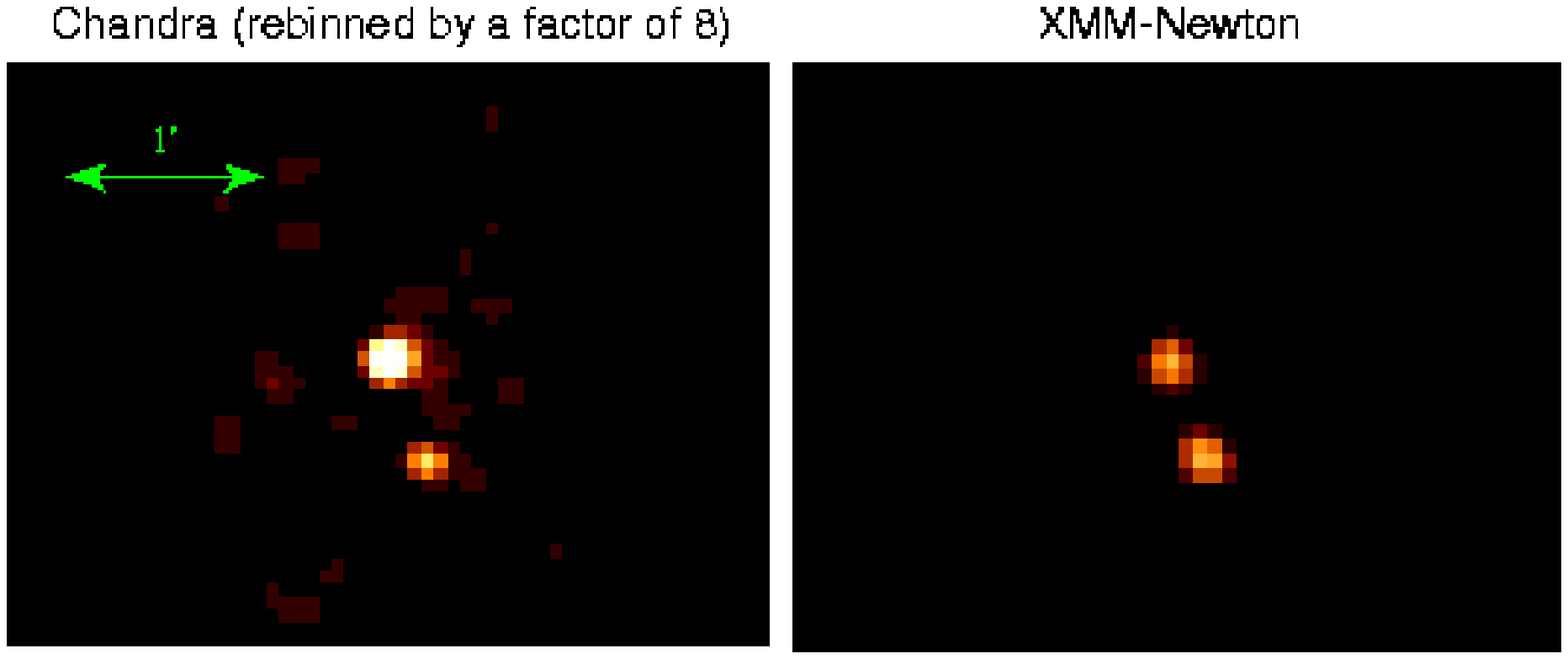,width=16cm}
\end{tabular}
\figcaption{The {\it Chandra}/ACIS-S (left; rebinned by a factor of
eight) and {\it XMM-Newton} (right; the MOS1, MOS2, and pn are
combined to produce this image) images of KS 1731--260. Both images
have been convolved with a Gaussian function with a width equal to the
image pixel sizes.  The source in the center is KS 1731--260 and the
other source to the right bottom is 2MASSI J173412.7--260548. The
images were scaled in such a way that the constant source 2MASSI
J173412.7--260548 has roughly the same appearance in both images, to
show the decrease in luminosity of KS 1731--260.
\label{fig:images} }
\end{center}
\end{figure}

We combined the data of the three EPIC cameras to create one image of
the field of KS 1731--260, representing the most sensitive image of
this region so far obtained. In Figure~\ref{fig:images}, we show both
the {\it Chandra}/ACIS-S (left) and the XMM-Newton/EPIC (right) images
of KS 1731--260. The {\it Chandra} image was rebinned by a linear
factor of 8 to obtain roughly the same pixel size as that of the {\it
XMM-Newton} image (3.95$''$ for the {\it Chandra} image vs. 4.35$''$
for the {\it XMM-Newton} image) and both images have been smoothed
using a Gaussian function with a width equal to the pixel size of the
image.  We clearly detected KS 1731--260 together with the nearby star
2MASSI J173412.7--260548 (Fig.~\ref{fig:images} right), both of which
were also detected during the {\it Chandra} observation
(Fig.~\ref{fig:images} left; Wijnands et al. 2001).  To allow for a
visual comparison, we used a scaling such that the appearance of this
2MASS star is very similar in both images (below we will show that the
flux of this star is consistent with being constant between the two
observations). A comparison of the images indicates that KS 1731--260
has decreased in luminosity between the {\it Chandra} and {\it
XMM-Newton} observations. In principle, systematic effects due to the
difference in the energy response of the instruments and the different
X-ray spectra of the two detected sources might be responsible for
this dimming of KS 1731--260 relative to the 2MASS star. However,
below we show that the decrease in luminosity as observed for KS
1731--260 is real.

\subsection{The source spectra}

The spectrum of KS 1731--260 in each EPIC camera was extracted using a
circle of 15$''$ in radius on the source position. The background
spectra were extracted from a circle with a radius of 50$''$ close to
KS 1731--260 (different background regions resulted in very similar
results) which did not contain any other point source (the standard
practice of using an annulus around the source position as background
could not be used because of the presence of the 2MASS source $\sim
30''$ away from KS 1731--260).  The extracted spectra were rebinned
using the FTOOLS routine GRPPHA into bins with a minimum of 10 counts
per bin.  We used the ready-made response matrices provided by the
calibration team (available at http://xmm.vilspa.esa.es/ccf/epic/).
We fitted the three spectra simultaneously using XSPEC version 11.1
(Arnaud 1996). We used several models to fit the data, and the neutral
hydrogen column density $N_{\rm H}$ was either fixed to
$1.1\times10^{22}$ cm$^{-2}$ (see, e.g., Barret et al. 1998 or Narita
et al. 2001) or left as a free parameter. All single-component models
resulted in acceptable fits. Currently, the two models most often used
to fit the quiescent spectra of neutron star systems are the blackbody
and the neutron star atmosphere models. Therefore, we concentrated on
those models, with the neutron star atmosphere model being that
described by Zavlin, Pavlov, \& Shibanov 1996 (the non-magnetic
case). In certain systems, a power-law tail above a few keV was found,
and although such power-law component was not required by the data, we
fitted the spectra with the above two models including a power-law
component with a photon index of 1 or 2 to obtain an upper limit on
this component.

\begin{figure}
\begin{center}
\begin{tabular}{c}
\psfig{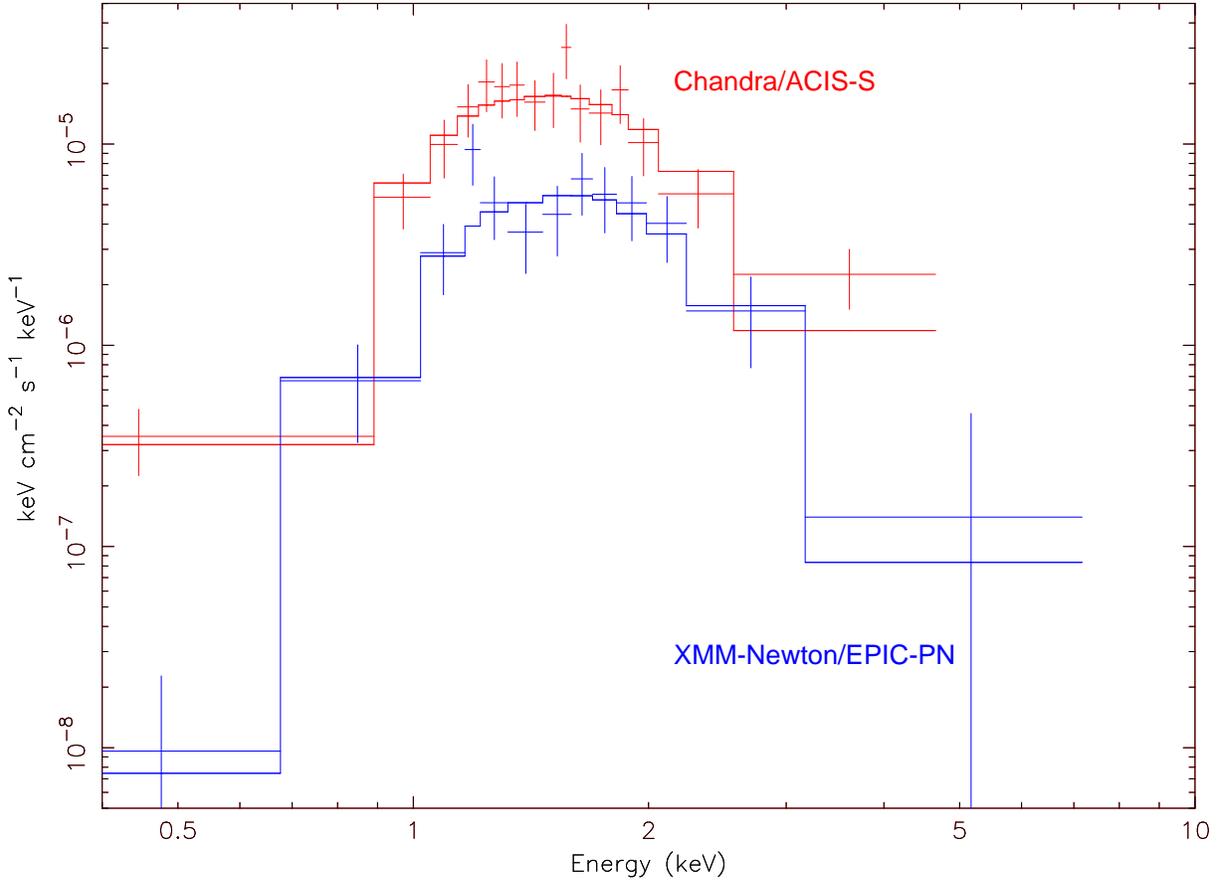}
\end{tabular}
\figcaption{The {\it Chandra}/ACIS-S (top spectrum; see also Wijnands
et al. 2001) and {\it XMM-Newton}/EPIC-pn (bottom) quiescent spectra
of KS 1731--260. The solid lines represent the best blackbody fits to
the data.
\label{fig:spectra} }
\end{center}
\end{figure}

The spectral results are listed in Tab.~\ref{tab:spectra} and the pn
spectrum is shown in Fig.~\ref{fig:spectra}. We have also plotted the
spectrum obtained with {\it Chandra} (Wijnands et al. 2001), which
again suggests that the source was fainter during our {\it XMM-Newton}
observation than during the {\it Chandra} observation.  When left
free, $N_{\rm H}$ was consistent with the value previously obtained
with other instruments, although for the atmosphere model a slightly
higher value was preferred, resulting in a slightly higher unabsorbed
flux compared to the fixed $N_{\rm H}$ case. When $N_{\rm H}$ was
fixed, the atmosphere model measured a similar flux as the blackbody
model, $\sim5\times10^{-14}$ \funit~(unabsorbed and for 0.5--10 keV).
To obtain the errors on the fluxes, we have calculated the $1\sigma$
error contours for the temperature and normalization, fixing $N_{\rm
H}$ at the value in Table~\ref{tab:spectra} in each case, and obtained
the fluxes associated with the circumference of the error ellipse.
The temperature $kT$ and $N_{\rm H}$ are strongly correlated in the
fits, and when both are free no useful constraints could be obtained
on the unabsorbed flux. The best-fit temperature was in all cases
$\sim$0.3 keV for the blackbody fits and $\sim$0.1 keV for the
atmosphere model. In the latter model, the neutron star radius could
not be constrained and was fixed to the best fit radius of 15 km (at
infinity; the other parameters are not very sensitive to its actual
value).  When including a power-law component in the fit, it could not
be detected significantly and its 0.5--10 keV flux could be
constrained to be less than 25\% of that obtained from the blackbody
or atmosphere component.

The images and spectra of KS 1731--260 both indicate that the flux
decreased between the two observation epochs. To investigate whether
the apparent flux decrease is statistically significant, we have
fitted the {\it Chandra} and {\it XMM-Newton} data
simultaneously. When all spectral parameters were tied between the two
data sets, a blackbody fit is statistically unacceptable, with
$\chi^2$ = 83 for 38 degrees of freedom, corresponding to a
probability of only $3 \times 10^{-5}$ that the source did not
change. We obtained a similar result when we used other models (e.g.,
atmosphere models) instead of a blackbody. When we did not tie the
spectral parameters (except $N_{\rm H}$ which was assumed to be
constant), we obtained acceptable fits. The fit results using a
blackbody or an atmosphere model are listed in
Table~\ref{tab:spectra}. In all cases, the flux difference between the
{\it Chandra} and {\it XMM-Newton} data is significant at a 3 to 4
$\sigma$ level.

Although this shows that the flux of KS 1731--260 decreased, this
could conceivably be due to a calibration error in one or both of the
instruments. Although this is unlikely (e.g., Ferrando et al. 2002;
Weisskopf et al. 2002), we can perform a check on this in the same
data set by analyzing the data of both instruments of the 2MASS star
assuming that the star has a constant spectrum.  To this end, we have
extracted the spectra of this source from the {\it Chandra} and {\it
XMM-Newton} data.  We fitted all obtained spectra of the 2MASS star
simultaneously keeping all spectral parameters tied between both
instruments (note that due to low statistics the {\it Chandra} data
alone did not allow to constrain the source spectrum). Either a
blackbody or a power-law spectrum fit the data well, yielding a
probability of only 0.12 (blackbody model) or 0.09 (power-law model)
that the flux of the 2MASS star changed by the same factor (a factor
of 3.5) as we observed for KS 1731--260. Furthermore, a recent cross
calibration study between {\it XMM-Newton} and {\it Chandra} (Snowden
2002) indicates that, for sources with different intrinsic spectra,
the measured fluxes of both instruments agree to within 10\%. Both
these results reinforce the idea that, the flux decrease we observed
in KS 1731--260 is real, and not due to calibration problems in any of
the two instruments.

\begin{figure}
\begin{center}
\begin{tabular}{c}
\psfig{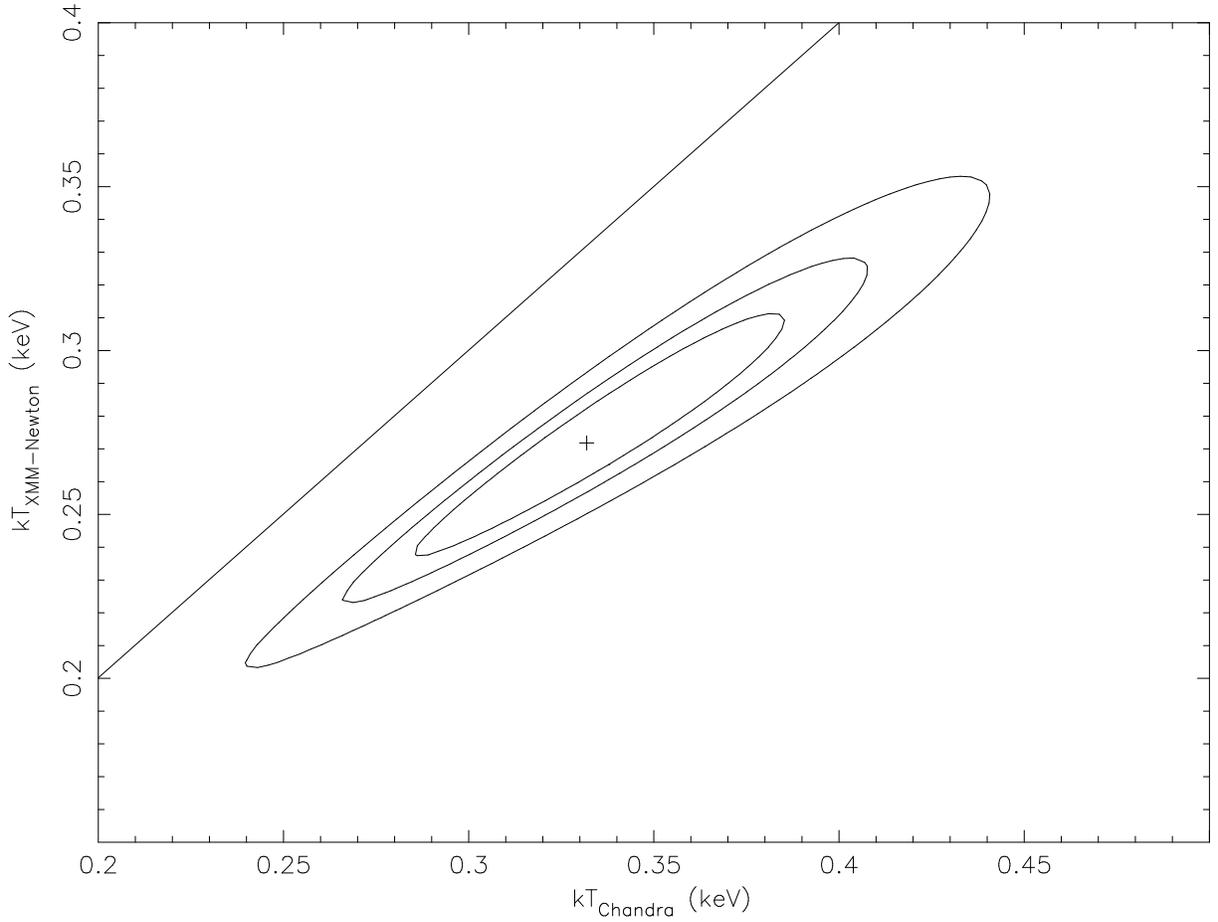}
\end{tabular}
\figcaption{The error ellipse of the {\it Chandra} temperature
$kT_{\rm Chandra}$ versus the {\it XMM-Newton} temperature $kT_{\rm
XMM-Newton}$. The model used in XSPEC was the bbodyrad model. The plus
represent the best fit value. The ellipses are for 1, 2, and 3$\sigma$
confidence level (for two parameters). The solid line is the line
$kT_{\rm Chandra} = kT_{\rm XMM-Newton}$.
\label{fig:temp} }
\end{center}
\end{figure}

In the cooling neutron star model, this flux decrease is due to a
temperature decrease. To investigate this, we fitted the two data sets
simultaneously with a blackbody model, letting $kT$ float between the
two observations, but keeping the same $N_{\rm H}$ and emitting
radius. This resulted in a $kT$ of $0.33^{+0.06}_{-0.05}$ and
0.27\pp0.04 keV for {\it Chandra} and {\it XMM-Newton},
respectively. The error ellipse of the two temperature parameters
(Fig.~\ref{fig:temp}) exclude the line $kT_{\rm Chandra} = kT_{\rm
XMM-Newton}$, which shows that in the constant-radius blackbody model
a systematically lower $kT$ is preferred to fit the {\it XMM-Newton}
spectrum than the {\it Chandra} one, suggesting that the temperature
decreased between the two epochs.  The resulting fluxes are 11\pp2
({\it Chandra}) and 4.8\pp0.8 $\times10^{-14}$ \funit ({\it
XMM-Newton}). This results in a flux decrease between the two data
sets of 6\pp2 $\times10^{-14}$ \funit, which is significant at the
3$\sigma$ level.

\section{Discussion\label{section:discussion}}

We have reported on {\it XMM-Newton} observations performed on 2001
September 13--14 of the neutron star X-ray transient KS 1731--260 when
it was in quiescence. We detected the source at an unabsorbed 0.5--10
keV flux of $\sim 4 - 8 \times 10^{-14}$ \funit, which for a distance
of 7 kpc implies a 0.5--10 keV luminosity of $\sim 2-5 \times 10^{32}$
\Lunit, depending on the model used to fit the data. This luminosity
is lower than what has been reported for the source during the {\it
Chandra} observation performed about half a year earlier (Wijnands et
al. 2001; Rutledge et al. 2002). KS 1731--260 is not the only system
for which X-ray variability in quiescence has been observed.  Several
other neutron star systems have also been found to be variable in
quiescence by factors of 3 to 5 on time scales of days to years (see
Ushomirsky \& Rutledge 2001 for a summary of the observed
variability).

It is expected that at some level the neutron star in KS 1731--260
should emit X-rays due to the thermal cooling of the neutron star
core.  Our low X-ray flux provides an upper limit to the thermal flux
from the core.  If the crust of the neutron star has a higher
temperature than the core (Rutledge et al. 2002 argued that the crust
should be considerably hotter than the core due to the prolonged
accretion episode of KS 1731--260) and/or if additional X-ray
production mechanisms are at work in the system (e.g., residual
accretion, radio pulsar mechanism), then the thermal flux related to
the core will be even lower.  Based on the Brown et al. (1998) model
and assuming standard core cooling, Wijnands et al. (2001) already
calculated that KS 1731--260 had to be in quiescence for over a 1000
years between outbursts in order to emit at the low flux level
measured with {\it Chandra} (see also Rutledge et al. 2002 or Burderi
et al. 2002). However, for the factor of 2 to 4 lower quiescent
luminosity we observed with {\it XMM-Newton}, this inferred cooling
time increases by approximately the same factor. This would make the
quiescent intervals of KS 1731--260 extremely long.  However, if we
assume that enhanced cooling takes place in the neutron star core
(e.g., due to enhanced neutrino production), this inferred quiescent
interval would decrease considerably, making it more similar to that
of the ordinary transients.

In the cooling neutron star model, the variability we observe would
have to be explained by assuming that the neutron star surface has
cooled between the {\it Chandra} and {\it XMM-Newton} observations. In
the previous section we have presented evidence that the measured
temperature decreased, supporting this interpretation.  For KS
1731--260, Rutledge et al. (2002) calculated four crust cooling curves
assuming different values of the crustal conductivity and the
different cooling processes in the core of the neutron star (standard
vs. enhanced cooling). A comparison of the observed decrease in
luminosity with those cooling curves (see Figure 3 in Rutledge et
al. 2002), suggests that our data are only consistent with a highly
conductive crust, and likely also enhanced core cooling occurs. For a
low heat conductivity in the crust, the X-ray luminosity of the system
should remain constant or even increase slightly, in contrast to what
we observed. For a highly conductive crust but only standard core
cooling a decrease in luminosity is also predicted, but by an amount
that is less than we have observed. However our uncertainties in the
actual luminosity decrease are considerable and our data might still
be consistent with this possibility. As explained above, the low
measured flux of the system by itself already suggests that enhanced
core cooling occurs.

In order to calculate the cooling curves, Rutledge et al. (2002)
assumed quiescent episodes for KS 1731--260 of 1500 years, which was
calculated assuming standard core cooling. However, if enhanced core
cooling occurs, the neutron star core can cool more rapidly than
assumed and the system could have quiescent episodes of only years to
decades.  This has to be taken into account in the modeling of the
cooling curves. Although the exact implications are unclear, our
conclusion that the crust has to be highly conductive to explain the
rapid cooling of the crust is unlikely to change.  Therefore, within
the cooling neutron star model, our new results indicate that the
neutron star in KS 1731--260 has a highly conductive crust and
enhanced cooling is likely to occur in its core.  Colpi et al. (2001)
suggested that when the mass of the neutron star exceeds $\sim1.6$
\mzon, such enhanced core cooling might occur. A massive neutron star
in KS 1731--260 is not unexpected because a significant amount of
matter must have been accreted in order for the neutron star to be
spinning rapidly. A fast spinning neutron star (with a spin frequency
of $\sim$524 Hz) in KS 1731--260 has been inferred from the burst
oscillations detected in this system (Smith, Morgan, \& Bradt 1997).

Alternative models explaining the quiescent emission in neutron star
X-ray transients have to be considered as well. In models assuming
that the emission is due to residual accretion onto the neutron star
surface, either directly or via leakage through the magnetospheric
barrier, or down to the magnetospheric radius, the detected luminosity
decrease can be explained by assuming that the accretion rate has
decreased considerably. These alternative models were discussed in
detail by Burderi et al. (2002) and because the lower luminosity of KS
1731--260 does not strongly affect their conclusions (an upper limit
on the magnetic field strength of the neutron star can be obtained
that is a factor $\sim$2 lower), we will not discuss those models here
in detail.  Note, that if the luminosity is indeed due to residual
accretion, the decrease in accretion rate inferred from our luminosity
decrease might cause a change in the X-ray production mechanism due to
the fact that the magnetospheric radius might move outside the
co-rotation radius or outside the light cylinder (see also Burderi et
al. 2002).

With further monitoring observations of KS 1731--260 in quiescence,
the quiescent properties of this source and their time evolution will
be better constrained. More detailed observations of other quiescent
neutron star systems will help to understand how similar they are to
KS 1731--260.  So far, at least two other systems have been identified
with similar properties: X 1732--304 (Wijnands, Heinke, \& Grindlay
2002) and 4U 2129+47 (Wijnands 2002; Nowak, Heinz, \& Begelman 2002).
Those systems also have very long outburst durations, and from their
quiescent properties it has been inferred that they should be in
quiescence for hundreds of years if only standard neutron star core
cooling occurs.  This spurred Wijnands et al. (2002) to suggest that
in the standard cooling scenario a correlation between the duration of
the outburst episodes and that of the quiescent intervals might be
required. However, such a correlation is difficult to understand in
accretion disk instability models (Lasota 2001). This could indicate
that enhanced cooling takes place in the neutron star cores of those
systems (Wijnands et al. 2002).  Our results indicating that enhanced
cooling may occur in the neutron star core of KS 1731--260 lends
further support to this idea.

\acknowledgments

We thank {\it XMM-Newton} project scientist Fred Jansen for scheduling
the observations used in this {\it Letter}.  RW was supported by NASA
through Chandra Postdoctoral Fellowship grant number PF9-10010 awarded
by CXC, which is operated by SAO for NASA under contract
NAS8-39073. This research has made use of the data and resources
obtained through the HEASARC online service, provided by NASA-GSFC.

\clearpage

\clearpage

\clearpage

\begin{deluxetable}{lcccc}
\tablecolumns{4}
\tablewidth{0pt} 
\tablecaption{Spectral results\label{tab:spectra}}
\tablehead{\multicolumn{5}{c}{{\it XMM-Newton} data}\\
Parameter& \multicolumn{2}{c}{Blackbody} & \multicolumn{2}{c}{Hydrogen atmosphere}}
\startdata
$N_{\rm H}$ ($10^{22}$ cm$^{-2}$) & $1.1^{+0.6}_{-0.4} $   & 1.1 (fixed)               &  $1.3^{+0.3}_{-0.4}$    & 1.1 (fixed)\\                           
$kT$ (keV)                        & $0.30^{+0.06}_{-0.05}$ & $0.30^{+0.04}_{-0.03}$    &  $0.11^{+0.03}_{-0.04}$ & $0.12^{+0.04}_{-0.02}$\\                
$F$                               & 4.8$^{+1.0}_{-0.9}$    & 4.8\pp1.0                 &  $7.4^{+2.3}_{-1.6}$    & $5.6^{+1.3}_{-1.1}$\\                               
$\chi^2$/dof                      & 15.7/23                & 15.8/24                   &  15.9/23                & 16.0/24\\                               
\hline
\multicolumn{5}{c}{Combined {\it Chandra} and {\it XMM-Newton} data} \\
Parameter& \multicolumn{2}{c}{Blackbody} & \multicolumn{2}{c}{Hydrogen atmosphere}\\
\hline
$N_{\rm H}$ ($10^{22}$ cm$^{-2}$) & $0.9^{+0.4}_{-0.3} $   & 1.1 (fixed)               & 1.0\pp0.2               &  1.1 (fixed) \\                         
$kT_{\rm XMM-N}$ (keV)            & 0.31\pp0.05            & $0.30^{+0.04}_{-0.03}$    & $0.14^{+0.04}_{-0.05}$  & $0.12^{+0.03}_{-0.02}$\\                
$kT_{\rm Chandra}$ (keV)          & $0.29^{+0.06}_{-0.05}$ & 0.27\pp0.03               & $0.11^{+0.03}_{-0.04}$  & 0.11\pp0.02           \\                
$F_{\rm XMM-N}$                   & $3.8^{+0.7}_{-0.5}$    & 5\pp1                     & $5.0^{+1.2}_{-0.8}$     & 6\pp1        \\                           
$F_{\rm Chandra}$                 & $13^{+3}_{-2}$         & $17^{+4}_{-3}$            & $18^{+4}_{-3}$          & $20^{+5}_{-4}$       \\                          
$\Delta F$                        & $9^{+3}_{-2}$          & $12^{+4}_{-3}$            & $13^{+4}_{-3}$          & $14^{+5}_{-4}$       \\                           
$\chi^2$/dof                      & 23.8/36                & 24.6/37                   & 23.2/36                 & 23.1/37       \\                               
\enddata
 
\tablenotetext{\,}{Note: The error bars represent 90\% confidence
levels, except for the fluxes for which the errors represent 68\%
confidence levels. The fluxes are unabsorbed, in the 0.5--10 keV
range, and in units of $10^{-14}$ \funit. For the blackbody we used
the bbodyrad model in XSPEC. For the hydrogen atmosphere model we used
the model of Zavlin et al. 1996 (the non-magnetic case) and for this
model the neutron star mass was fixed to 1.4 \mzon, its radius at
infinity to 15 km, the distance was assumed to be 7 kpc, and the
temperature is for an observer at infinity}

\end{deluxetable}

\end{document}